\documentclass[useAMS]{mn2e}
 \usepackage{times}
 \usepackage{graphicx}
 \usepackage{epsfig}
 \usepackage{amssymb}

 \title[Reconstructing the Chelyabinsk event]
       {Reconstructing the Chelyabinsk event: pre-impact orbital evolution}

 \author[C. de la Fuente Marcos and R. de la Fuente Marcos]
        {C.~de~la~Fuente~Marcos\thanks{E-mail: nbplanet@fis.ucm.es}
         and
         R. de la Fuente Marcos \\
         Universidad Complutense de Madrid,
         Ciudad Universitaria, E-28040 Madrid, Spain}
 \date{Accepted 2014 May 25.
       Received 2014 April 7;
       in original form 2013 November 15}
 \pubyear{2014}
 
\begin{document}
  \maketitle

  \begin{abstract}
     The Chelyabinsk superbolide was the largest known natural object to enter 
     the Earth's atmosphere since the Tunguska event in 1908 and it has become a 
     template to understand, manage and mitigate future impacts. Although the 
     event has been documented in great detail, the actual pre-impact orbit of 
     the parent body is still controversial. Here, we revisit this topic using 
     an improved Monte Carlo approach that includes the coordinates of the 
     impact point to compute the most probable solution for the pre-impact orbit 
     ($a$~=~1.62 au, $e$~=~0.53, $i$~=~3$\fdg$97, $\Omega$~=~326$\fdg$45 and 
     $\omega$~=~109$\fdg$71). We also check all the published solutions using a
     simple yet robust statistical test to show that many of them have problems 
     to cause an impact at the right time. We use the improved orbit and 
     $N$-body simulations to revisit the dynamical status of a putative 
     Chelyabinsk asteroid family and confirm that it could be linked to resonant 
     asteroids 2007 BD$_{7}$ and 2011~EO$_{40}$. In addition, and as the 
     classification of Chelyabinsk meteorites is well established, a search for 
     meteorite falls of the same chondrite group and petrologic type gives some 
     evidence for the existence of an associated LL5 chondrite cluster. 
  \end{abstract}


  \section{Introduction}
     An asteroidal impact of the magnitude of the Chelyabinsk event has never before been documented in so much detail. Yet it happened 
     during daytime and that makes the analysis of the available observational material much harder. There is full consensus among the 
     scientific community on the dynamical class of the impactor, it was an Apollo asteroid. Also widely accepted is that it was a piece of 
     a larger object and that additional fragments may be following similar orbits (Borovi\v{c}ka et al. 2013; de la Fuente Marcos \& de la 
     Fuente Marcos 2013, hereafter Paper I; Popova et al. 2013). However, the actual values of its orbital parameters are still a matter of 
     debate and different authors have published solutions for which the overall dispersions in range are relatively large (see Table 
     \ref{orbits}): 31 per cent in semimajor axis, 27 per cent in eccentricity, 49 per cent in inclination, 0.09 per cent in longitude of 
     the node and 24 per cent in argument of perihelion. This translates into multiple theories on possible links between the parent body 
     and certain near-Earth asteroids (NEAs) or groups of them: 86039 (1999 NC$_{43}$) (Borovi\v{c}ka et al. 2013), the Flora family 
     (Itokawa, Popova et al. 2013) or 2011 EO$_{40}$ and related objects in Paper I. 

     Here, we revisit the topic of the pre-impact orbit of the Chelyabinsk superbolide using an improved Monte Carlo methodology that 
     includes the coordinates of the impact point. This Letter is organized as follows. The current status of the subject of the calculation 
     of the pre-impact orbit is summarized in Section 2; a statistically robust impact test aimed at validating candidate orbits is 
     outlined and applied to the published solutions, results are discussed. An improved, most probable pre-impact orbit is found in 
     Section 3. Using the improved orbit and $N$-body simulations we study the dynamical status of a putative Chelyabinsk asteroid family in 
     Section 4. In Section 5 we introduce some additional supporting evidence in the form of a meteorite cluster analysis. Results are 
     briefly discussed and conclusions summarized in Section 6.
%
%
     \begin{table*}
      \centering
      \fontsize{8}{11pt}\selectfont
      \tabcolsep 0.05truecm
      \caption{Published solutions for the pre-impact orbit of the Chelyabinsk superbolide tested here. The errors from Popova et al. (2013) 
               are from their Table S5B. Paper I is de la Fuente Marcos \& de la Fuente Marcos (2013). The values of the probabilities 
               discussed in the text are also given.}
      \begin{tabular}{lccccccccc}
       \hline
          Authors                     & $a$ (au)          & $e$               & $i$ (\degr)      & $\Omega$ (\degr)      & $\omega$ (\degr)      & 
          $P_{\rm 0.050 au}$ & $P_{r_{\rm H}}$ & $P_{10 R_{\rm E}}$ & $n_{\rm c}$ \\
       \hline
          Borovi\v{c}ka et al.        & 1.55$\pm$0.07     & 0.50$\pm$0.02     & 3.6$\pm$0.7      & 326.410$\pm$0.005     & 109.7$\pm$1.8         &
          0.889              & 0.231           & 0.009    & 440  \\
          Borovi\v{c}ka et al. (2013) & 1.72$\pm$0.02     & 0.571$\pm$0.006   & 4.98$\pm$0.12    & 326.459$\pm$0.001     & 107.67$\pm$0.17       &
          1                  & 0.716           & 0.032    & 0    \\
          Paper I                     & 1.62375$\pm$0.00014 & 0.53279$\pm$0.00011 & 3.817$\pm$0.005  & 326.4090$\pm$0.0007   & 109.44$\pm$0.03       &
          1                  & 1               & 0.897    & 99499 \\
          Nakano$^5$                  & 1.6223665$\pm$0.0000001 & 0.5311191$\pm$0.0000001 & 3.87128$\pm$0.00001 & 326.42524$\pm$0.00001           & 109.70844$\pm$0.00001           &
          1                  & 1               & 0.920    & 546405 \\
          Popova et al. (2013)        & 1.76$\pm$0.04     & 0.581$\pm$0.009   & 4.93$\pm$0.24    & 326.4422$\pm$0.0014   & 108.3$\pm$1.9         &
          1                  & 0.412           & 0.015    & 35   \\
          Proud (2013)                & 1.47$^{+0.03}_{-0.13}$ & 0.52$^{+0.01}_{-0.05}$ & 4.61$^{+2.58}_{-2.09}$ & 326.53$^{+0.01}_{-0.0}$ & 96.58$^{+2.94}_{-1.72}$ &
          0.768              & 0.183           & 0.006    & 0    \\
          Zuluaga \& Ferrin$^6$       & 1.73$\pm$0.23     & 0.51$\pm$0.08     & 3.45$\pm$2.02    & 326.70$\pm$0.79       & 120.62$\pm$2.77       &
          0.341              & 0.071           & 0.001    & 15   \\
          Zuluaga et al. (2013)       & 1.27$\pm$0.05     & 0.44$\pm$0.02     & 3.0$\pm$0.2      & 326.54$\pm$0.08       & 95.1$\pm$0.8          &
          0.245              & 0.068           & 0.002    & 0    \\
          Zuluaga et al.$^7$          & 1.368$\pm$0.006   & 0.470$\pm$0.010   & 4.0$\pm$0.3      & 326.479$\pm$0.003     & 99.6$\pm$1.3          &
          0.250              & 0.173           & 0.007    & 0    \\
       \hline
          \,\,\,\, average$\pm\sigma$ & 1.6$\pm$0.2       & 0.52$\pm$0.04     & 4.0$\pm$0.7      & 326.49$\pm$0.09       & 106$\pm$8             &
                             &                 &          &      \\
       \hline
          This work                   & 1.624765$\pm$0.000005 & 0.53184$\pm$0.00001 & 3.97421$\pm$0.00005  & 326.44535$\pm$0.00001   & 109.71442$\pm$0.00004       &
          1                  & 1               & 0.919    & 148765  \\
       \hline
      \end{tabular}
      \label{orbits}
     \end{table*}
%
%

  \section{The pre-impact orbit so far}
     The dynamical class of the Chelyabinsk impactor has been well established since the beginning, calculations by Adamo (2013), 
     Borovi\v{c}ka et al. (2013), Borovi\v{c}ka et al. (Green 2013), Chodas \& Chesley,\footnote{http://neo.jpl.nasa.gov/news/fireball\_130301.html} 
     Emel'yanenco et al.,\footnote{http://www.inasan.ru/eng/asteroid\_hazard/chelyabinsk\_bolid\_new.html}
     Lyytinen,\footnote{http://www.amsmeteors.org/2013/02/large-daytime-fireball-hits-russia/} Lyytinen, Matson \& 
     Gray\footnote{http://www.projectpluto.com/temp/chelyab.htm}, Nakano,\footnote{http://www.icq.eps.harvard.edu/CHELYABINSK.HTML}
     Popova et al. (2013), Proud (2013), Zuluaga \& Ferrin\footnote{http://arxiv.org/abs/1302.5377}, Zuluaga, Ferrin \& Geens
     (2013),\footnote{http://astronomia.udea.edu.co/chelyabinsk-meteoroid/} and Paper I showed that the parent body was an Apollo asteroid. 
     In contrast, and despite the multiplicity of published orbital solutions, the actual pre-impact orbit of the parent body of the 
     Chelyabinsk superbolide (see Table \ref{orbits}, the rest can be found in Paper I) is still controversial. The overall ranges for the 
     orbital elements are in some cases too large for comfort (see above). Any computed orbit must be consistent with an obvious fact, on 
     2013 February 15, 03:20:33 GMT a superbolide was observed in the skies near Chelyabinsk, Russia. 

     The orbital elements and therefore the position of our planet at the time of the impact are well known (see Table \ref{Earth} and 
     Appendix A). Any acceptable orbital determination must put the parent body in the immediate neighbourhood of the Earth on that day and 
     time. The orbital parameters of any given solution are characterized by errors and any analysis of the validity of a given orbit must 
     be discussed in statistical terms. If, for a certain solution, the probability of being close to the Earth at the impact time is below 
     a reasonable threshold, the orbit must be rejected. Under the two-body approximation, the equations of the orbit of an object around 
     the Sun in space are given by the expressions (e.g. Murray \& Dermott 1999): 
     \begin{eqnarray}
        X & = & r \ (\cos \Omega \cos(\omega + f) - \sin \Omega \sin(\omega + f) \cos i)
                    \nonumber \\
        Y & = & r \ (\sin \Omega \cos(\omega + f) + \cos \Omega \sin(\omega + f) \cos i)
                    \label{orbit} \\
        Z & = & r \ \sin(\omega + f) \sin i \nonumber
     \end{eqnarray}
     where $r = a (1 - e^{2})/(1 + e \cos f)$, $a$ is the semimajor axis, $e$ is the eccentricity, $i$ is the inclination, $\Omega$ is the
     longitude of the ascending node, $\omega$ is the argument of perihelion and $f$ is the true anomaly. Using the above equations and the
     data in Table \ref{Earth}, the position of the Earth is uniquely determined. For a given orbit with known ranges for the orbital 
     parameters (given by the standard deviations or errors) the impact risk assessment can be performed by means of a Monte Carlo 
     simulation (Metropolis \& Ulam 1949). 

\begin{figure*}
        \centering
        \includegraphics[width=5.7cm]{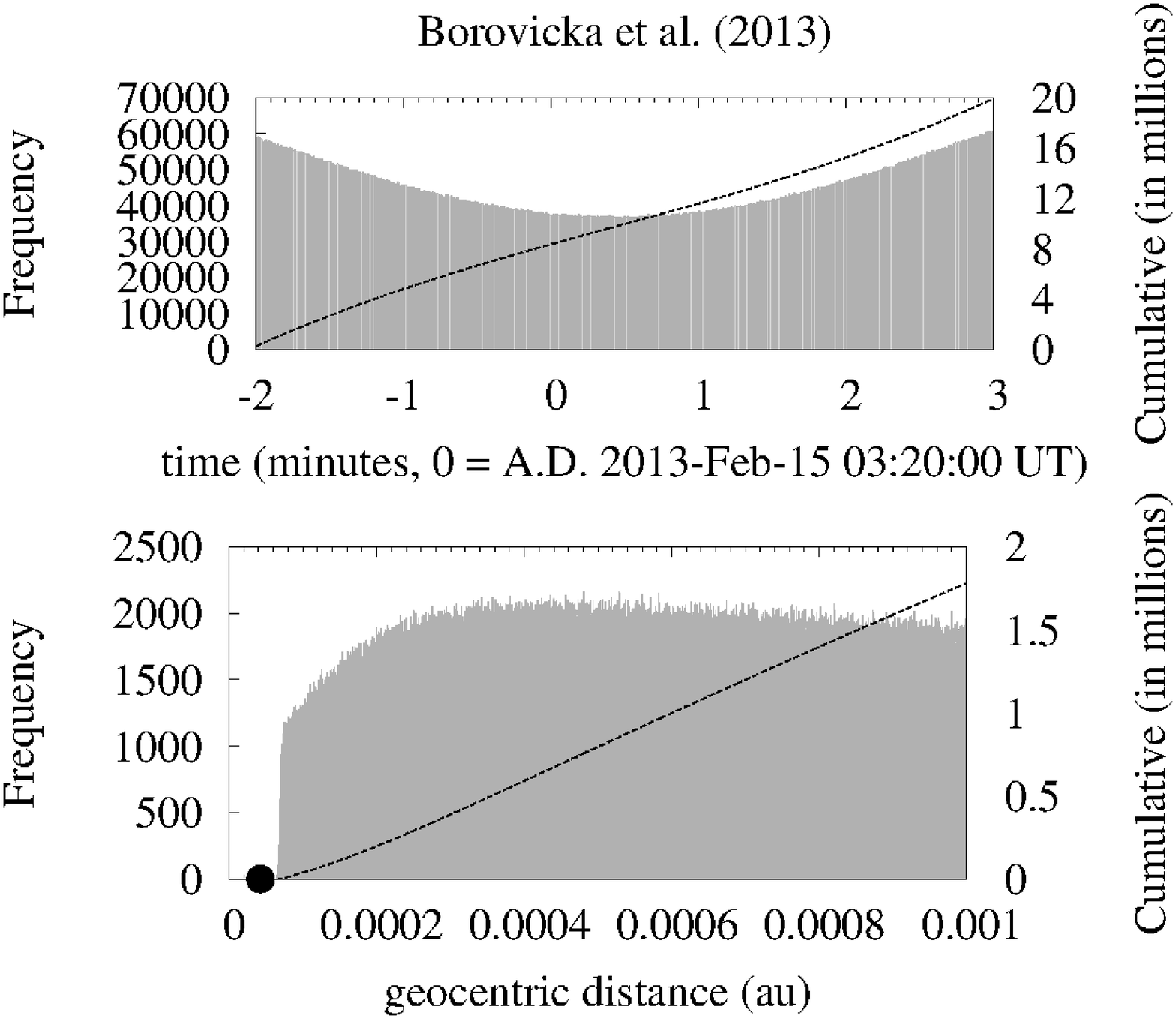}
        \includegraphics[width=5.7cm]{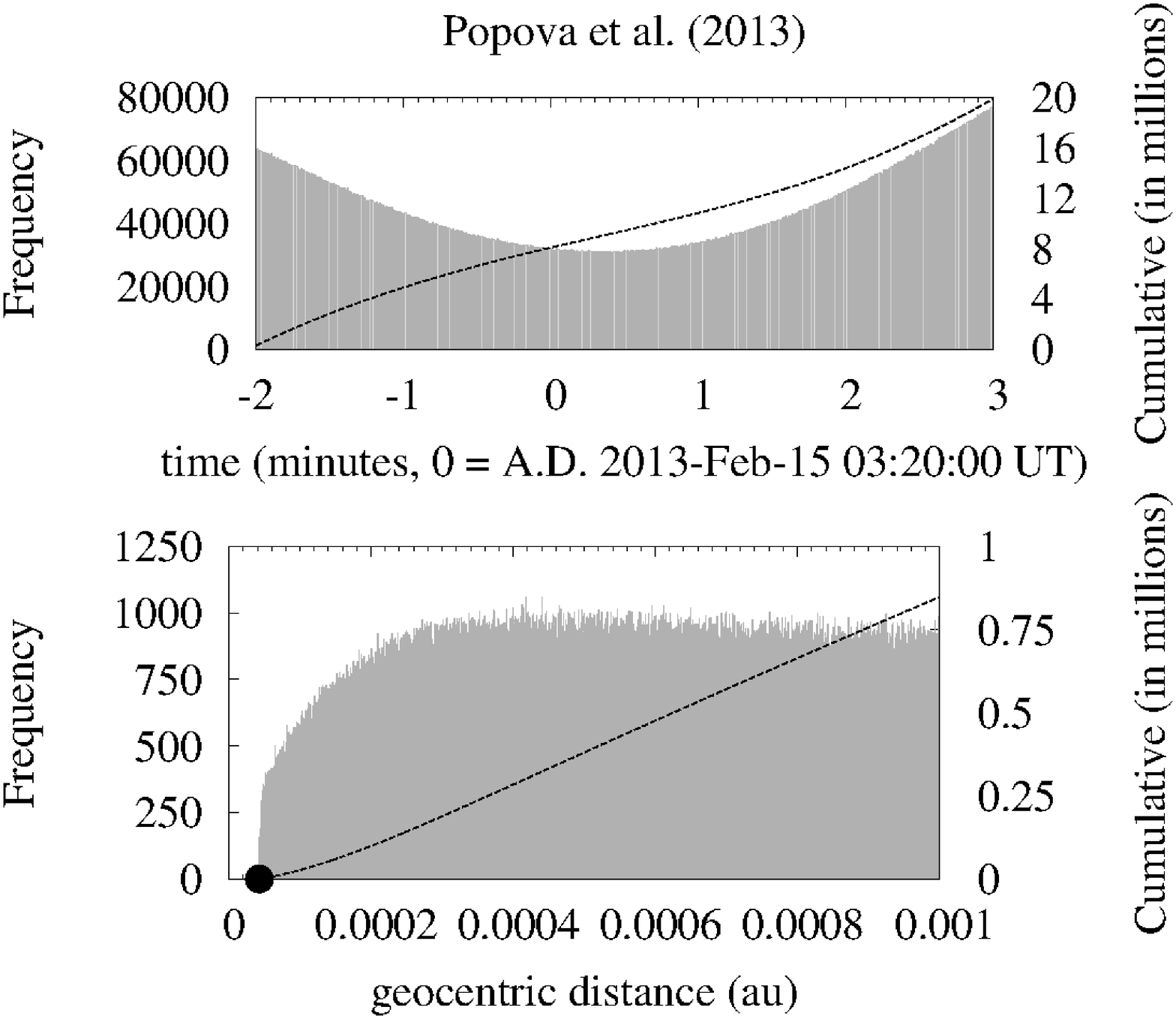}
        \includegraphics[width=5.7cm]{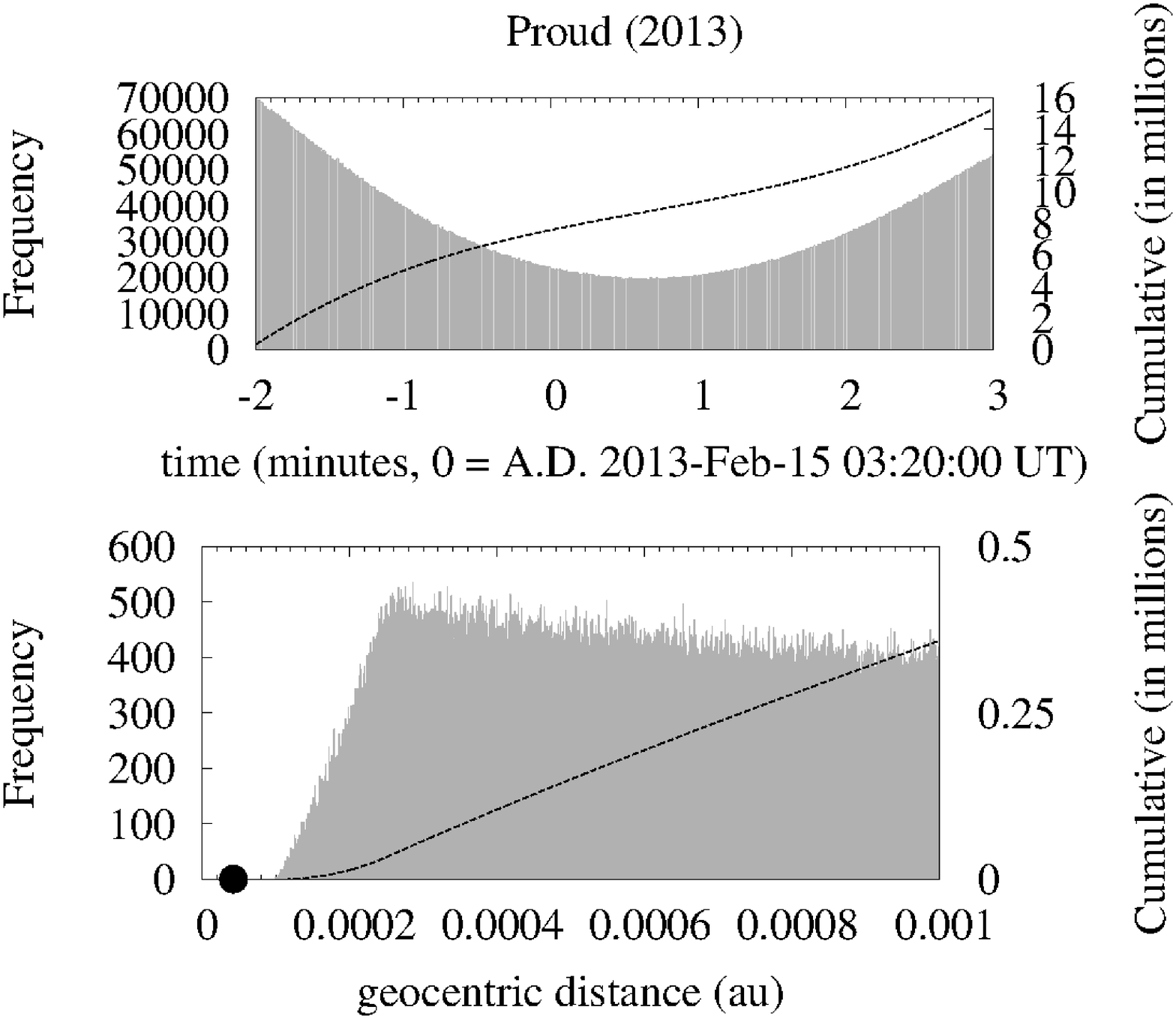}\\
        \includegraphics[width=5.7cm]{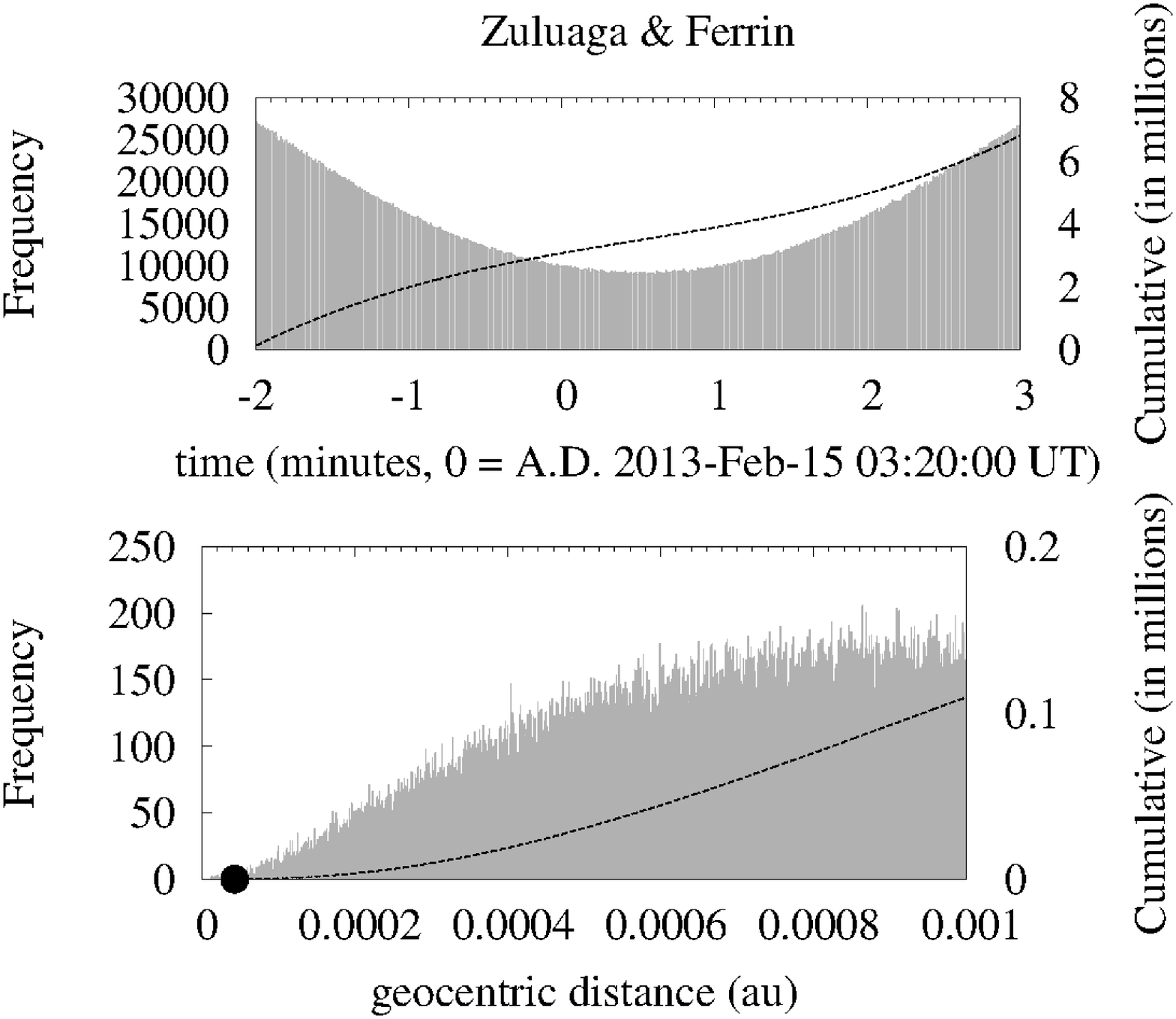}
        \includegraphics[width=5.7cm]{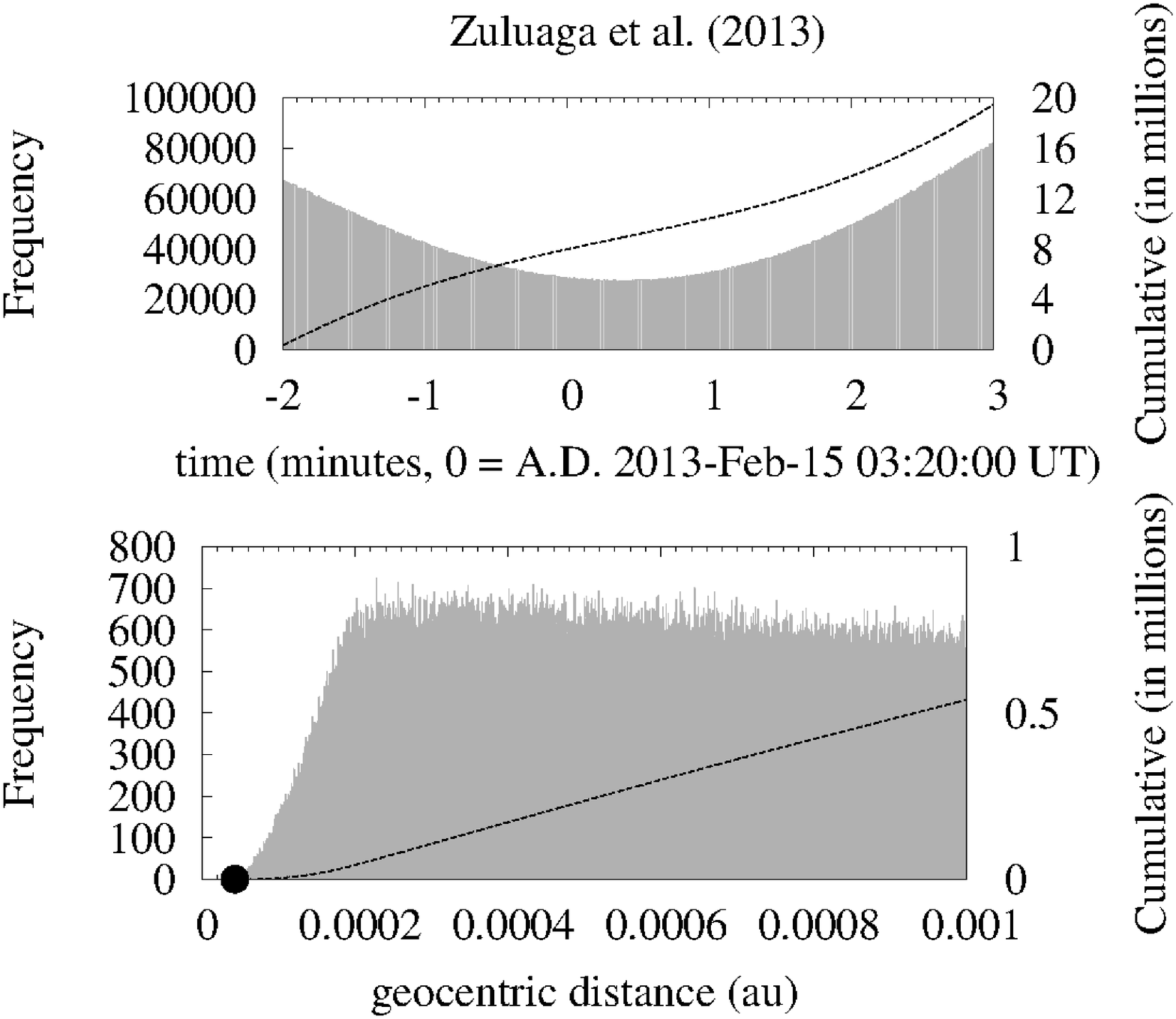}
        \includegraphics[width=5.7cm]{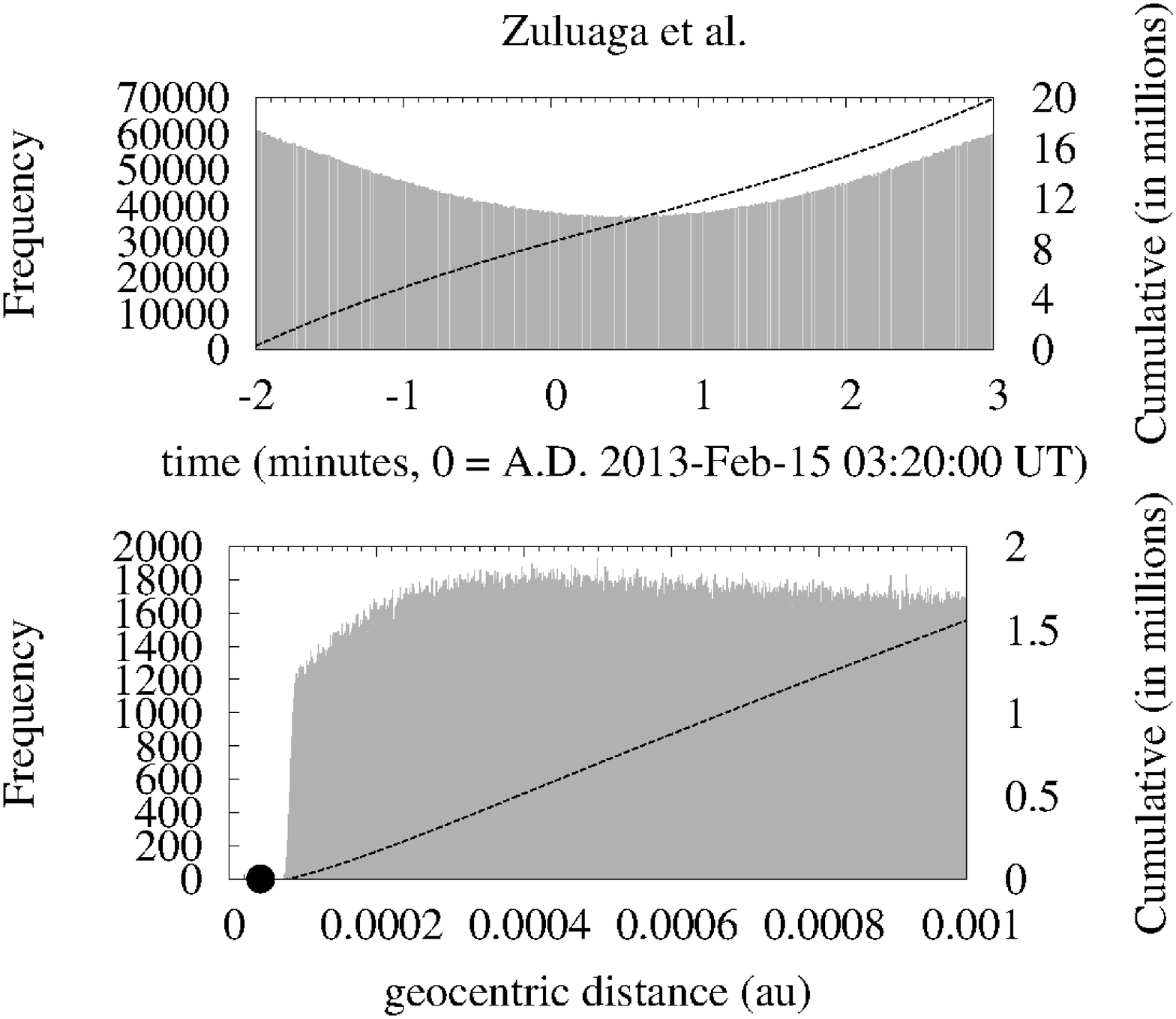}\\
        \includegraphics[width=5.7cm]{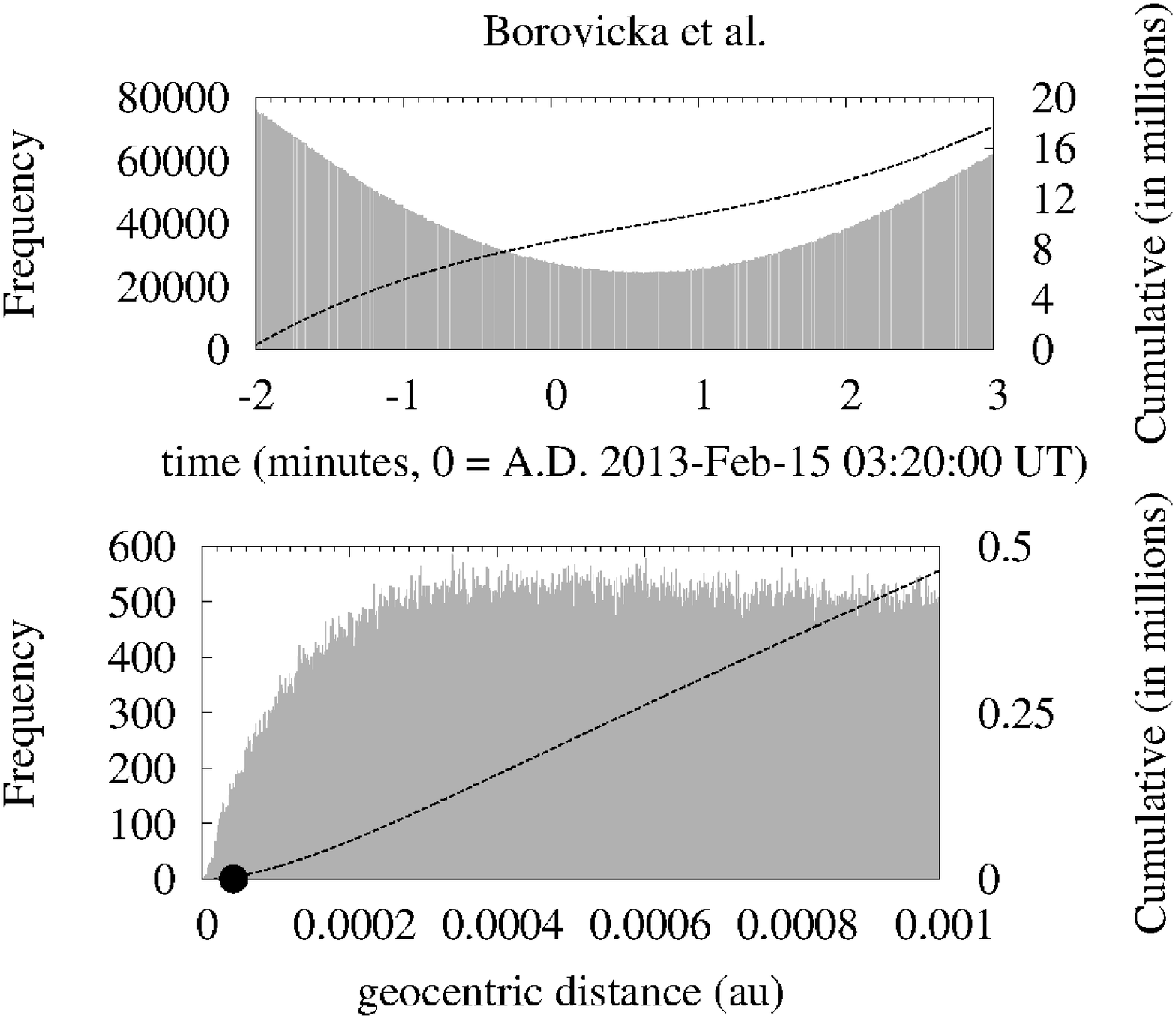}
        \includegraphics[width=5.7cm]{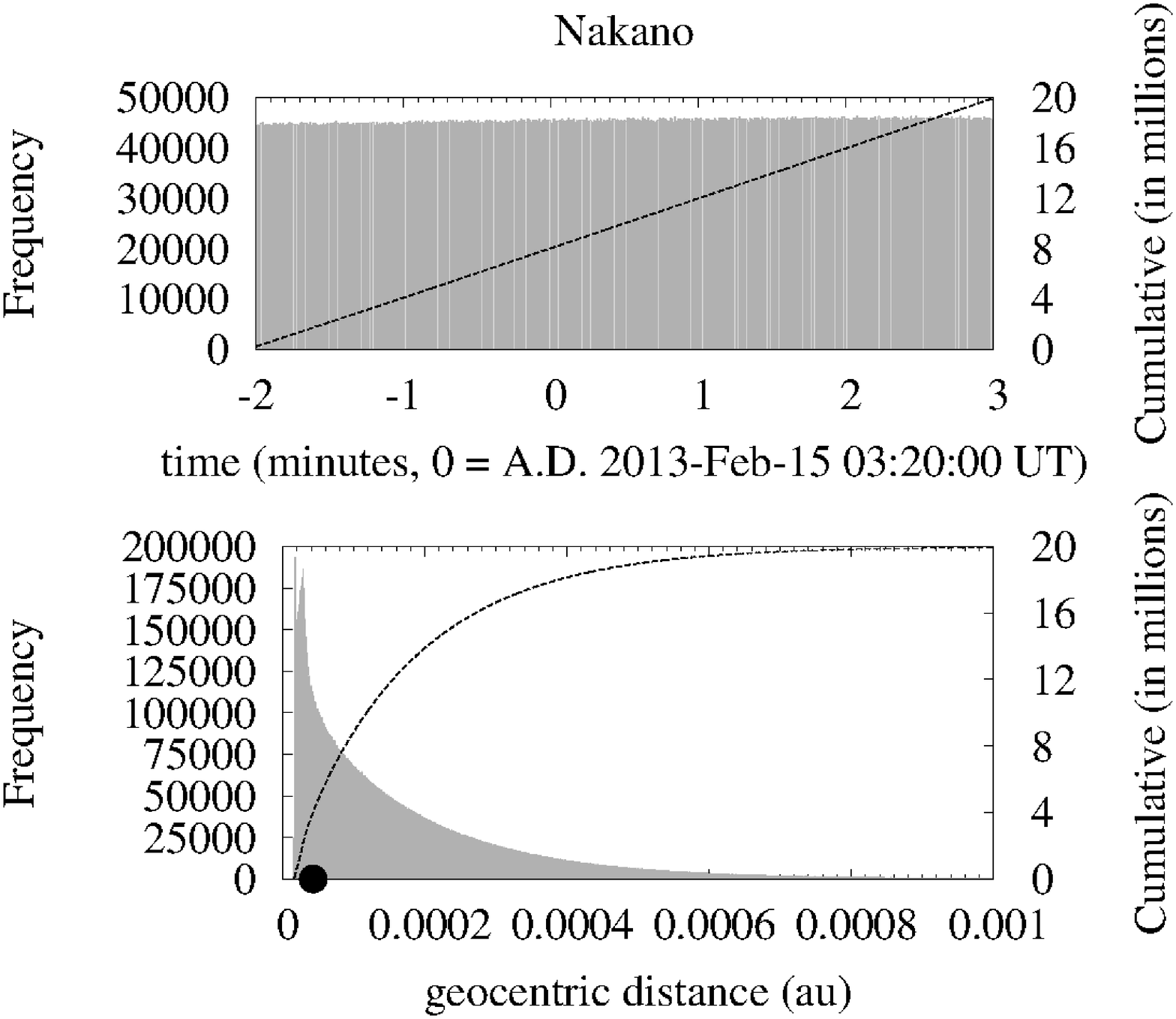}
        \includegraphics[width=5.7cm]{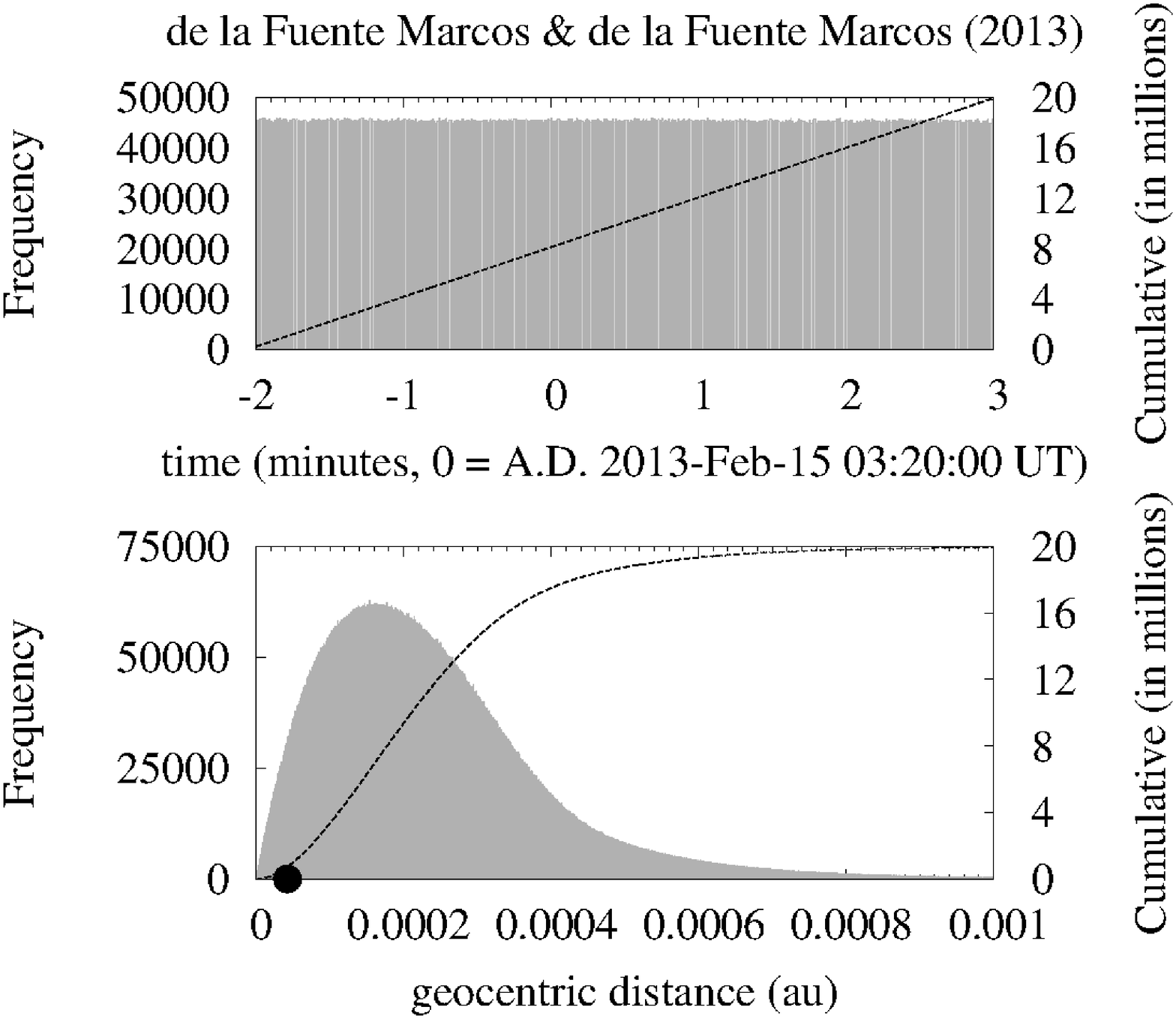}\\
        \caption{Distribution in time and geocentric distance of the MOIDs associated with the solutions in Table \ref{orbits}. The time is 
                 referred to JDCT 2456338.6389 = A.D. 2013-Feb-15 03:20:00. The black dot represents the radius of the Earth. Here and in 
                 Fig. \ref{ours} the bin size in time is 0.001 min and 0.000001 au in distance. 
                 } 
        \label{stats}
     \end{figure*}

     Let us consider a set of orbital elements ($a$, $e$, $i$, $\Omega$ and $\omega$) for the incoming body. These elements are randomly 
     sampled within fixed (assumed) ranges following a uniform distribution. For each set, we randomly sample the above equations in true 
     anomaly for both the object and the Earth, computing the usual Euclidean distance between both points so the minimal distance is 
     eventually found. This value coincides with the minimum orbit intersection distance (MOID) used in Solar system studies. The assumed 
     range for the true anomaly of the Earth is small, equivalent to a time interval of about 6 min, approximately centred at the impact 
     time. In comparison, the time taken by our planet to travel a distance equal to its own average diameter (12\,742 km, $R_{\rm E}$ = 
     6\,371 km) is nearly 7.1 minutes. For an object following an impact trajectory, the largest orbital changes take place when it is 
     within 10 Earth radii ($R_{\rm E}$) from the Earth's centre (see e.g. Jenniskens et al. 2009; Oszkiewicz et al. 2012). If the predicted 
     perigee (MOID) of an object is mostly (in probabilistic terms) outside 10 $R_{\rm E}$, the actual probability of impact is negligible. 
     Any candidate solution predicting a perigee for the parent body beyond 0.000425 au (10\ $R_{\rm E}$) should be rejected. Our 
     7-dimensional Monte Carlo sampling provides a robust statistical impact validation for any input orbit. If we apply the algorithm 
     described above to the solutions in Table \ref{orbits} we obtain Fig. \ref{stats}. There, we show the distribution in time and 
     geocentric distance for the MOIDs associated with the various solutions. A total of 2$\times10^7$ test orbits have been computed for 
     each solution. The probabilities of having a MOID under 0.05 au ($P_{\rm 0.050 au}$) and 0.0004263 au (10\ $R_{\rm E}$, $P_{10 R_{\rm E}}$) 
     at the time of impact are given in Table \ref{orbits}. Only two orbits have a probability higher than 50 per cent of placing the 
     impactor within 10\ $R_{\rm E}$ of our planet at impact time. For objects following the other orbits, the chances of being near the 
     Earth at impact time are smaller, in some cases significantly. Our simple, yet robust statistical test makes just one single and very 
     reasonable assumption: that the data in Table \ref{Earth} are correct. It may be argued that our impact test is based on the two-body 
     approximation but the impactor penetrates well inside a region where that approximation is no longer valid because the gravitational 
     field of the Earth, not the Sun, is dominant. However, if we focus on a sphere centred on the Earth and of radius equal to the Hill 
     radius of our planet ($r_{\rm H}$ = 0.0098 au) which is the conventional limit for its sphere of influence, we still find problems with 
     most solutions. Any serious candidate impact solution must have a probability ($P_{r_{\rm H}}$) close to 1 for the MOIDs to be under 
     one Hill radius around impact time. Only two solutions in Table \ref{orbits} fulfil that condition. Besides, focusing on the number of 
     orbits reaching a distance to the surface of the Earth $<$ 100 km (the characteristic thickness of the Earth's atmosphere) within 20 
     seconds of JDCT 2456338.6391296, $n_c$, the results are also consistent (see Table \ref{orbits}). However, no data on the impact point 
     were used to compute the solution in Paper I. 
%
%
%
%

  \section{Improved Monte Carlo analysis}
\begin{figure}
        \centering
        \includegraphics[width=\linewidth]{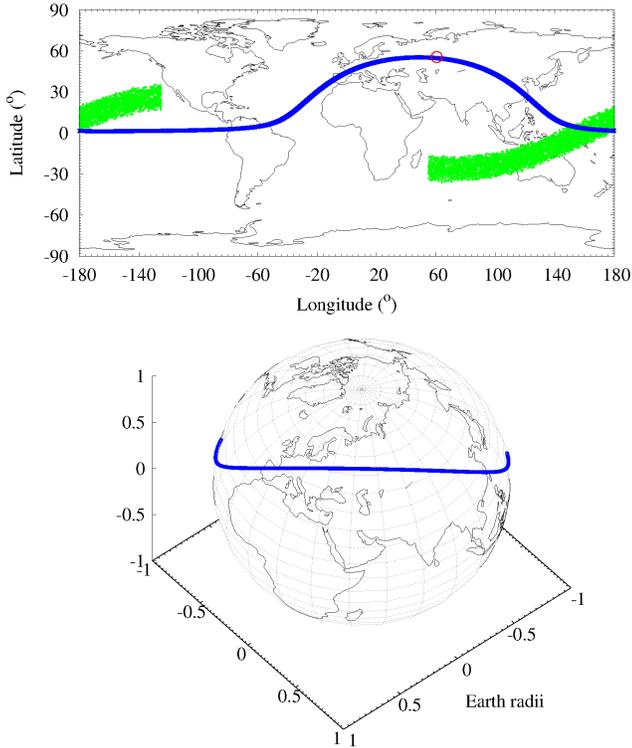}
        \caption{Path of risk for one representative orbit obtained with our improved Monte Carlo analysis; it reached an altitude over the 
                 surface of the Earth of 97.26 km at coordinates (63\fdg90~E, 54\fdg48~N). The blue curve outlines the flight path (E to W) 
                 of the object assuming that it did not hit the ground. The green stripe represents the impact risk associated with the 
                 solution in Paper I. The location of the city of Chelyabinsk is also plotted.
                }
        \label{chelyabinsk}
     \end{figure}

\begin{figure}
        \centering
        \includegraphics[width=\linewidth]{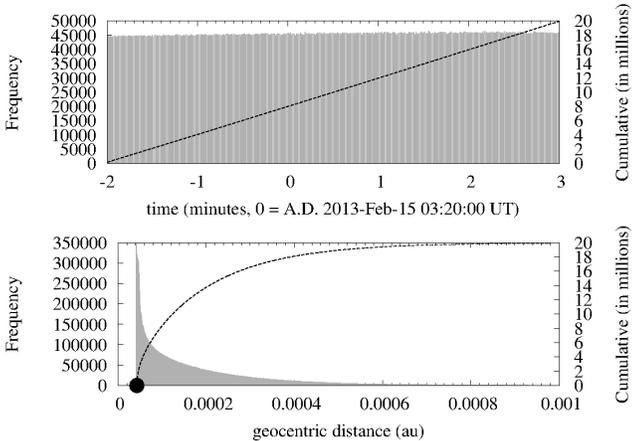}
        \caption{As Fig. \ref{stats} but for our preferred solution (see Table \ref{orbits}).
                 } 
        \label{ours}
     \end{figure}

     In Paper I we made a simplifying assumption by considering a point-like Earth, neglecting the geographical coordinates of the impact 
     point (60.32\degr~E, 54.96\degr~N), a hole on the frozen surface of Chebarkul Lake (Popova et al. 2013). As a result, all the impact 
     points associated with that solution are near the Earth's equator (green points in Fig. \ref{chelyabinsk}), not close to the city of 
     Chelyabinsk. To further improve our previous orbit we computed the coordinates of the impact point for our test orbits as described in 
     e.g. Montenbruck \& Pfleger (2000). In computing the longitude of impact, we assume that the MOID happens when the object is directly 
     overhead (is crossing the local meridian). Under that approximation, the local sidereal time corresponds to the right ascension of the 
     object and its declination is the latitude of impact. Instead of using as impact point the hole in Chebarkul Lake, we use the location 
     of the actual atmospheric entry. The superbolide was first detected on 2013-Feb-15 03:20:20.8$\pm$0.1 s UT at longitude 
     64\fdg565$\pm$0\fdg030, latitude +54\fdg445$\pm$0\fdg018 and altitude 97.1$\pm$0.7 km (see Table S1, Popova et al. 2013, also Miller et 
     al. 2013). The new most probable orbit is not too different from the one in Paper I and, again, matches well the one originally 
     computed by S. Nakano (see Table \ref{orbits}). It was found after about 10$^{10}$ trials. The distribution in time and geocentric 
     distance of the MOIDs associated with our preferred solution (see Fig. \ref{ours}) as well as the various probabilities in Table 
     \ref{orbits} show that this orbit is statistically more robust than any of the published solutions (except Nakano's). Figure 
     \ref{chelyabinsk} displays its associated path of risk. The particular orbit depicted there (see Table \ref{impactor}) had an altitude 
     of over 97 km at coordinates (63\fdg9~E, 54\fdg5~N) and reached perihelion early on 2012 December 31. Geometrically, it is the most 
     probable orbit. Using the coordinates of the hole in Chebarkul Lake gives similar orbital solutions but we believe that our choice is 
     technically more correct.  
%
%
%
%
%

%
%
%
%

  \section{Related objects and dynamical evolution}
     Assuming that the object responsible for the Chelyabinsk event was a fragment of a larger body (or that other objects move in similar 
     orbits), we use the D-criteria of Southworth \& Hawkins (1963), $D_{\rm SH}$, Lindblad \& Southworth (1971), $D_{\rm LS}$, Drummond 
     (1981), $D_{\rm D}$, and the $D_{\rm R}$ from Valsecchi, Jopek \& Froeschl\'e (1999) to investigate possible dynamical connections 
     between this object and known minor bodies. A search among all the objects currently catalogued (as of 2014 April 7) by the Jet 
     Propulsion Laboratory (JPL) Small-Body Database\footnote{http://ssd.jpl.nasa.gov/sbdb.cgi} using these criteria gives the list of 
     candidates in Table \ref{candidatesCH}. With one exception, their orbits are poorly known as they are based on short arcs. All of them 
     are classified as Apollos, NEAs and, a few, as potentially hazardous asteroids (PHAs); their aphelia are in or near the 3:1 orbital 
     resonance with Jupiter (at 2.5 au). These objects are strongly perturbed as they experience periodic close encounters not only with the 
     Earth--Moon system but also with Mars, Ceres and, in some cases, Venus. They are also submitted to multiple secular resonances (see 
     below). We have studied the short-term past and future orbital evolution of several of these objects using the Hermite integration 
     scheme described by Makino (1991) and implemented by Aarseth (2003). Our physical model includes the perturbations by the eight major 
     planets, the Moon, the barycentre of the Pluto-Charon system and the five largest asteroids. For accurate initial positions and 
     velocities, we used the elements provided by the JPL online Solar system data service\footnote{http://ssd.jpl.nasa.gov/?planet\_pos} 
     (Giorgini et al. 1996) and based on the DE405 planetary orbital ephemerides (Standish 1998) referred to the barycentre of the Solar 
     system. For more details see de la Fuente Marcos \& de la Fuente Marcos (2012) and Paper I. 

\begin{figure}
        \centering
        \includegraphics[width=\linewidth]{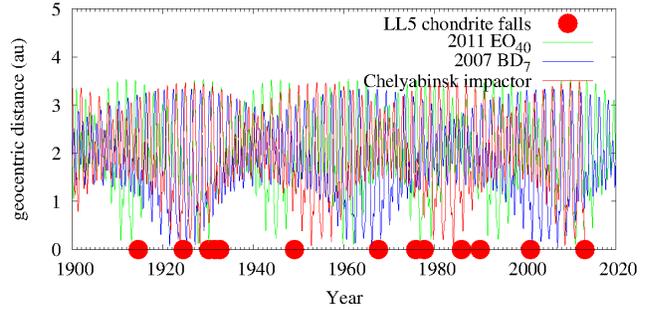}
        \caption{Distance from the Earth to the Chelyabinsk impactor, 2007 BD$_{7}$ and 2011~EO$_{40}$ since 1900. Observed LL5 chondrite
                 falls are also indicated (see Table \ref{meteorites}). Note the periodic flybys (see the text for details).  
                 }
        \label{resonance}
     \end{figure}

     Figure \ref{resonance} shows that the pre-impact orbit of the Chelyabinsk superbolide was also affected by multiple secular resonances. 
     The object was experiencing one or more close encounters with the Earth-Moon system at nearly 27 yr intervals. Asteroid 2007 BD$_{7}$ 
     follows a similarly recurrent encounter sequence but this time with a period close to 44 yr. Asteroid 2011 EO$_{40}$ (and 1996 AW$_1$) 
     also undergoes close approaches to the Earth-Moon system following a rather regular pattern, every 17 yr approximately.  
%
%
%
%

  \section{An LL5 chondrite cluster?}
\begin{figure}
        \centering
        \includegraphics[width=\linewidth]{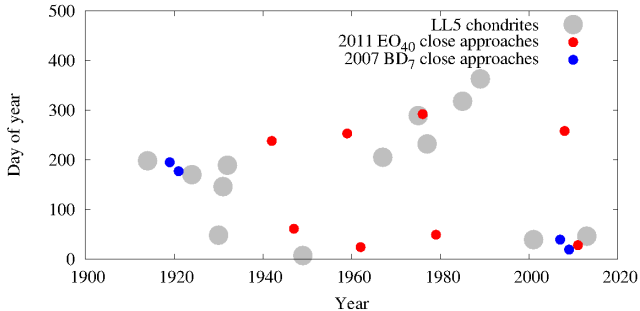}
        \caption{Year-day distribution of LL5 chondrite falls and flybys with asteroids 2007 BD$_{7}$ and 2011~EO$_{40}$ since 1900. Day one 
                 = January 1. Sources: see Table \ref{meteorites}, JPL Small-Body Database and NEODyS system.    
                 }
        \label{falls}
     \end{figure}

     It has been suggested that the time of fall of meteorites reflects the orbital distribution of their parent bodies. If the parent 
     bodies of certain meteorites are organized in orbital groups, fragments may collide with the Earth in different years but their 
     calendar dates should exhibit some linear correlation. Studying the existence of meteorites of the same chondrite group and petrologic 
     type that fell within some calendar days of each other in different years may uncover the existence of an orbital group or meteoroid
     stream. Figures \ref{resonance} and \ref{falls} appear to suggest such a trend (see also Table \ref{meteorites}). Encounters at the 
     descending node happen in January/February and encounters at the ascending node occur in June to August. A Plavchan periodogram 
     (Plavchan et al. 2008) gives a most significant period for the LL5 chondrite falls of 17.37 yr with a probability that the detected 
     value is due to chance of 0.0013. The second most significant period is 28.32 yr with a $p$-value of 0.0083. This is close to the 
     proposed flyby pattern period of the Chelyabinsk superbolide parent body with our planet. Although the evidence is certainly 
     encouraging, we must take these results with caution because the number of observed falls is small. 
%
%
%
%

  \section{Discussion and conclusions}
     In this Letter, we have obtained a statistically sound, most probable solution for the pre-impact orbit of the Chelyabinsk superbolide
     and implemented a simple yet robust Monte Carlo-based probability test to validate candidate solutions. The past dynamical evolution of 
     this most probable orbit is almost a textbook example of how meteorites are delivered to the Earth from the 3:1 orbital resonance with 
     Jupiter (at 2.5 au) as described by e.g. Gladman et al. (1997), see panel C in their Fig. 3. Therefore, the ultimate origin of the 
     Chelyabinsk superbolide can be tracked backwards to the main asteroid belt. The orbit of the parent body of the Paragould meteorite 
     (another LL5 chondrite) has also been traced back to the 3:1 resonance (Nelson \& Thomsen 1947). This resonance often pushes minor 
     bodies into the Sun or out of the resonance, towards the inner Solar system, creating transient near-Earth objects. For objects 
     following this evolutionary path and delivered to the region close to the semimajor axis range 1.5--1.6 au, the $\nu_3$, $\nu_4$ and 
     $2 g = g_5 + g_6$ secular resonances are dominant (Gladman et al. 1996; Michel \& Froeschl\'e 1997). This translates into horizontal 
     oscillations in the ($e$, $a$) plane (see Fig. \ref{horizontal}) as the secular resonances modify $e$ at constant $a$. The impactor 
     appears to be a dynamical relative of equally resonant asteroids 2007 BD$_{7}$ and 2011~EO$_{40}$. The dynamical relationship is 
     certainly encouraging but the current orbits of these asteroids are not reliable enough to claim a conclusive connection; a genetic 
     link in the form of a mutually consistent chondritic constitution (for the Chelyabinsk meteorites) and the asteroids' surface 
     composition remains to be tested. In the absence of a genetic link, this group of asteroids is still a family but a resonant one. Their 
     current orbital evolution results in a series of periodic close encounters with the Earth-Moon system (at 17 yr intervals for 
     2011 EO$_{40}$-like orbits, 27 yr for the Chelyabinsk impactor, or 44 yr for 2007 BD$_{7}$-like orbits), due to the combined action of 
     multiple secular resonances. One effect of this peculiar dynamical behaviour is that if one of these objects happens to be in an 
     adverse position for observing from the Earth at the time of its close approach, then the object will remain unobserved for decades. 
     Reaching perigee at small solar elongations makes these objects inherently difficult to discover and track from the ground. This is 
     exactly what happened with the Chelyabinsk impactor. Space-based observations are the only proper way to study this population. However, 
     during their flybys, they are good targets for study by Earth-based radar. If there are additional objects moving in similar orbits, 
     they may have struck the Earth in the past. Records of meteorite falls of the same chondrite group and petrologic type (LL5) provide 
     some marginal yet consistent evidence in favour of this scenario. A reasonably compatible match between the compositions and physical 
     properties of the Chelyabinsk meteoritical samples and those of any of the other meteorites in Table \ref{meteorites} (see e.g. 
     Olivenza versus Chelyabinsk) will provide definitive support for the analysis completed here.  

  \section*{Acknowledgements}
     The authors would like to thank the referee, P. Wiegert, for his reports, S. Nakano and S. J. Aarseth for their comments on an earlier 
     version of this Letter and S. J. Aarseth for providing one of the codes used in this research. This work was partially supported by the 
     Spanish `Comunidad de Madrid' under grant CAM S2009/ESP-1496. We thank M. J. Fern\'andez-Figueroa, M. Rego Fern\'andez and the 
     Department of Astrophysics of the Universidad Complutense de Madrid (UCM) for providing computing facilities. Most of the calculations 
     and part of the data analysis were completed on the `Servidor Central de C\'alculo' of the UCM and we thank S. Cano Als\'ua for his 
     help during this stage. In preparation of this Letter, we made use of the NASA Astrophysics Data System, the ASTRO-PH e-print server, 
     the MPC data server and the NEODyS system.

  \newpage
  \appendix
  \section{Orbital elements of the Earth around the time of impact}
     The superbolide was first detected on 2013-Feb-15 03:20:20.8$\pm$0.1 s UT at longitude 64\fdg565$\pm$0\fdg030, latitude 
     54\fdg445$\pm$0\fdg018 and altitude 97.1$\pm$0.7 km (see Table S1, Popova et al. 2013). Therefore, the actual impact with the 
     atmosphere took place at epoch 2456338.6391296 Julian Date, Coordinate Time. The geometric osculating orbital elements of the Earth 
     within approximately $\pm$150 s of the first detection are given in Table \ref{Earth}. These values have been computed by the Solar 
     System Dynamics Group, Horizons On-Line Ephemeris System. The time resolution provided by this ephemeris system is one minute and we 
     decided to use JDCT 2456338.6389 = A.D. 2013-Feb-15 03:20:00 as reference; this instant is considered as our $t = 0$ across this work 
     unless explicitly stated. Therefore, we assume that the entry of the superbolide started approximately 20.8 s after $t = 0$. The 
     orbital elements at that time (record in bold in Table \ref{Earth}) have been obtained by interpolation using the data in Table 
     \ref{Earth}.
%
%
     \begin{table*}
      \centering
      \fontsize{8}{11pt}\selectfont
      \tabcolsep 0.06truecm
      \caption{Orbital elements of the Earth around JDCT 2456338.6389 = A.D. 2013-Feb-15 03:20:00 (Source: JPL \textsc{Horizons} system).
               Data as of 2014 April 7.}
      \begin{tabular}{cccccccc}
       \hline
          Epoch JD CT       & CT         &
          $a$ (au)          & $e$                 & $i$ (\degr)          & $\Omega$ (\degr)  & $\omega$ (\degr)  & $f$ (\degr)       \\
       \hline
          2456338.637500000 & 03:18:00.0 &
          1.000460470380130 & 0.01681075152088699 & 0.003426162422992692 & 163.1611470947406 & 301.5002044683560 & 41.76083874450404 \\
          2456338.638194445 & 03:19:00.0 &
          1.000460358050270 & 0.01681064764891284 & 0.003426287278598604 & 163.1605191659505 & 301.5006652233060 & 41.76170745517870 \\
          2456338.638888889 & 03:20:00.0 &
          1.000460245713456 & 0.01681054377780261 & 0.003426412091192330 & 163.1598915292676 & 301.5011256430874 & 41.76257620868673 \\
      {\bf 2456338.6391296} & {\bf 03:20:20.8} &
         {\bf 1.0004602068} &  {\bf 0.0168105078} &   {\bf 0.0034264553} & {\bf 163.1596744389} & {\bf 301.5012846926} & {\bf 41.7628774485} \\
          2456338.639583333 & 03:21:00.0 &
          1.000460133369692 & 0.01681043990755927 & 0.003426536860777701 & 163.1592641843954 & 301.5015857279981 & 41.76344500502645 \\
          2456338.640277778 & 03:22:00.0 &
          1.000460021018978 & 0.01681033603818453 & 0.003426661587351488 & 163.1586371314587 & 301.5020454779188 & 41.76431384419267 \\
          2456338.640972222 & 03:23:00.0 &
          1.000459908661319 & 0.01681023216968093 & 0.003426786270908627 & 163.1580103702289 & 301.5025048930830 & 41.76518272618031 \\
       \hline
      \end{tabular}
      \label{Earth}
     \end{table*}
%
%

  \section{Supplementary materials}
%
%
     \begin{table*}
      \centering
      \fontsize{8}{11pt}\selectfont
      \tabcolsep 0.06truecm
      \caption{Representative orbital elements of the parent body of the Chelyabinsk superbolide around the time of impact.}
      \begin{tabular}{cccccccc}
       \hline
          Epoch JD CT       & CT         &
          $a$ (au)          & $e$                 & $i$ (\degr)          & $\Omega$ (\degr)  & $\omega$ (\degr)  & $M$ (\degr)       \\
       \hline
          2456338.6391296   & 03:20:20.8 &
          1.62476552        & 0.53184298          & 3.9742124            & 326.445352        & 109.714428        & 21.9204136        \\
       \hline
      \end{tabular}
      \label{impactor}
     \end{table*}
%
%
%
%
     \begin{table*}
      \centering
      \fontsize{8}{11pt}\selectfont
      \tabcolsep 0.07truecm
      \caption{Orbital elements, orbital periods ($P_{\rm orb}$), perihelia ($q = a \ (1 - e)$), aphelia ($Q = a \ (1 + e)$), number of 
               observations ($n$), data-arc, and absolute magnitudes ($H$) of the candidates to be the parent body of the meteoroid that 
               caused the Chelyabinsk superbolide. The various $D$-criteria ($D_{\rm SH}$, $D_{\rm LS}$, $D_{\rm D}$ and $D_{\rm R}$) are 
               also shown. The objects are sorted by ascending $D_{\rm R}$. Only objects with $D_{\rm R} < 0.05$ are shown. Data as of 2014 
               April 7.} 
      \begin{tabular}{ccccccccccccccccc}
       \hline
          Asteroid       & $a$ (au)  & $e$        & $i$ (\degr) & $\Omega$ (\degr) & $\omega$ (\degr) & $P_{\rm orb}$ (yr) & $q$ (au) & $Q$ (au) 
                         & $n$ & arc (d) & $H$ (mag) 
                         & $D_{\rm SH}$ & $D_{\rm LS}$ & $D_{\rm D}$ & $D_{\rm R}$ & PHA \\
       \hline
         2011 EO$_{40}$  & 1.6541021 & 0.54021638 & 3.36308     &  50.30833        &  17.06892        & 2.13               & 0.76     & 2.55
                         & 20  & 34      & 21.50 
                         & 0.1198       & 0.0136       & 0.0396      & 0.0073      & Yes \\
         2011 GP$_{28}$  & 1.5913963 & 0.51988326 & 4.04802     &  16.39708        & 252.18927        & 2.01               & 0.76     & 2.42
                         & 14  & 1       & 29.40 
                         & 1.0474       & 0.0125       & 0.4904      & 0.0096      & No  \\
         2002 AC$_{9}$   & 1.7037203 & 0.56057681 & 2.28439     &   2.59440        &  28.42421        & 2.22               & 0.75     & 2.66
                         & 51  & 3132    & 21.00 
                         & 0.4231       & 0.0429       & 0.1407      & 0.0225      & Yes \\
         2012 QZ$_{16}$  & 1.5380029 & 0.50339326 & 6.12109     & 151.62989        & 258.93207        & 1.91               & 0.76     & 2.31
                         & 23  & 2       & 25.50 
                         & 0.2902       & 0.0471       & 0.1007      & 0.0237      & No  \\
         2012 VA$_{20}$  & 1.6839686 & 0.55558020 & 4.39572     &  62.72875        & 240.13973        & 2.19               & 0.75     & 2.62
                         & 16  & 10      & 22.80 
                         & 1.0052       & 0.0277       & 0.4055      & 0.0260      & No  \\
         2013 BR$_{15}$  & 1.5543524 & 0.52032940 & 1.95427     & 102.89583        & 284.89541        & 1.94               & 0.75     & 2.36
                         & 10  & 2       & 25.00 
                         & 0.4426       & 0.0400       & 0.1464      & 0.0261      & No  \\
         2009 SD         & 1.7327530 & 0.56634130 & 3.04578     & 344.32656        & 287.05641        & 2.28               & 0.75     & 2.71
                         & 24  & 3       & 25.40 
                         & 1.0894       & 0.0392       & 0.5038      & 0.0261      & No  \\
         2008 UM$_{1}$   & 1.7553337 & 0.56564555 & 4.66126     & 208.94985        & 110.66782        & 2.33               & 0.76     & 2.75
                         & 8   & 1       & 32.10 
                         & 0.9422       & 0.0359       & 0.3575      & 0.0278      & No  \\
         1996 VB$_{3}$   & 1.6269482 & 0.54485464 & 2.79698     & 180.59557        & 132.69687        & 2.08               & 0.74     & 2.51
                         & 21  & 9       & 22.40 
                         & 0.9524       & 0.0316       & 0.3689      & 0.0294      & No  \\
         2014 AF$_{5}$   & 1.5675221 & 0.51935896 & 6.41544     & 100.66155        & 288.73603        & 1.96               & 0.75     & 2.38
                         & 24  & 1       & 28.80
                         & 0.4511       & 0.0450       & 0.1496      & 0.0301      & No  \\  
         2010 DU$_{1}$   & 1.6865098 & 0.53929611 & 3.70444     & 147.83186        &  74.25038        & 2.19               & 0.78     & 2.60
                         & 22  & 4       & 26.50 
                         & 1.0330       & 0.0186       & 0.4340      & 0.0336      & No  \\
         2013 UX         & 1.6994450 & 0.56015033 & 5.45431     & 259.60985        &  51.62281        & 2.21               & 0.75     & 2.65
                         & 53  & 42      & 22.00 
                         & 0.9725       & 0.0405       & 0.3800      & 0.0336      & No  \\
         2004 RN$_{251}$ & 1.6554455 & 0.52788736 & 4.39167     & 179.61123        & 245.93586        & 2.13               & 0.78     & 2.53
                         & 27  & 2       & 26.10 
                         & 0.1716       & 0.0225       & 0.0608      & 0.0356      & No  \\
         2008 EF$_{32}$  & 1.6264297 & 0.52172394 & 1.73277     & 349.17117        & 112.28006        & 2.07               & 0.78     & 2.47
                         & 8   & 1       & 29.40 
                         & 0.2353       & 0.0439       & 0.0768      & 0.0379      & No  \\
         2007 BD$_{7}$   & 1.5622989 & 0.49806111 & 4.84907     & 343.62635        & 219.85947        & 1.95               & 0.78     & 2.34
                         & 185 & 14      & 21.10 
                         & 0.9241       & 0.0439       & 0.3660      & 0.0383      & Yes \\
         2011 CZ$_{3}$   & 1.5962287 & 0.51076590 & 2.11312     & 326.23527        & 241.70608        & 2.02               & 0.78     & 2.41
                         & 30  & 4       & 26.30 
                         & 0.9527       & 0.0437       & 0.3825      & 0.0426      & No  \\
         2008 UT$_{95}$  & 1.8148434 & 0.57460717 & 3.81217     & 220.05056        & 247.41560        & 2.44               & 0.77     & 2.86
                         & 32  & 2       & 27.40 
                         & 0.3219       & 0.0443       & 0.1115      & 0.0447      & No  \\
         2008 FH         & 1.5821116 & 0.50363499 & 3.45434     &   5.20697        & 264.09093        & 1.99               & 0.79     & 2.38
                         & 25  & 12      & 24.30 
                         & 1.0304       & 0.0386       & 0.4814      & 0.0460      & No  \\
         1996 AW$_{1}$   & 1.5277410 & 0.51711311 & 4.73110     & 117.91581        & 228.70205        & 1.89               & 0.74     & 2.32
                         & 13  & 13      & 19.40 
                         & 0.7543       & 0.0303       & 0.2670      & 0.0487      & Yes  \\
       \hline
      \end{tabular}
      \label{candidatesCH}
     \end{table*}
%
%
%
%
     \begin{table*}
      \centering
      \fontsize{8}{11pt}\selectfont
      \tabcolsep 0.1truecm
      \caption{All the meteorites in this table have recommended classification LL5 and their falls were observed.
               Source: Meteoritical Bulletin Database unless otherwise indicated (http://www.lpi.usra.edu/meteor/metbull.php). 
               $^1$    Gismelseed A. M., Bashir S., Worthing M. A., Yousif A. A., Elzain M. E., Al Rawas A. S., Widatallah H. M., 2005, 
                       Meteoritics \& Planetary Science, 20, 255. 
               $^2$    Reed S. J. B., Chinner G. A., 1995,
                       Meteoritics, 30, 468. 
               $^3$    Wagner C., Arnold G., Wasch R., 1988,
                       Meteoritics, 23, 93. 
               $^4$    Al-Bassam K. S., 1978,
                       Meteoritics, 13, 257. 
               $^5$    Graham A. L., Michel-Levy M. C., Danon J., Easton A. J., 1988,
                       Meteoritics, 23, 321. 
               $^6$    Levi-Donati G. R., Sighinolfi G. P., 1974,
                       Meteoritics, 9, 1.
               $^{7}$  Mason B., Wiik H. B., 1964,
                       Geochimica et Cosmochimica Acta, 28, 533.
               $^{8}$  Mu\~noz-Espadas M. J., 2003, 
                       PhD thesis, Universidad Complutense de Madrid, Madrid, Spain. 
               $^{9}$  Meszaros M., Ditr\'oi-Pusk\'as Z., V\'aczi T., Kereszturi \'A., 2013,
                       44th Lunar and Planetary Science Conference, p.\ 1477.
               $^{10}$ Popova et al. (2013).
               }
      \begin{tabular}{cccccc}
       \hline
          Name                       &  Date fell       & Fayalite       & Ferrosilite    & Wollastonite   & $\epsilon^{53}$ Cr $^{10}$  \\
                                     &                  & (mol per cent) & (mol per cent) & (mol per cent) &                            \\
       \hline
          Chelyabinsk                & 2013 February 15 & 27.9$\pm$0.4/29.2$\pm$0.3$^{10}$ & 22.8$\pm$0.8   & 1.3$\pm$0.3    & 0.23$\pm$0.03              \\
          Al Zarnkh$^1$              & 2001 February 8  & 28             & 23             &                &                            \\
          Bawku$^2$                  & 1989 December 29 & 26.8           & 22.6           &                &                            \\
          Salzwedel$^3$              & 1985 November 14 & 26.8           & 23.9           &                &                            \\
          Alta'ameem$^4$             & 1977 August 20   & 27             & 24.5           &                &                            \\
          Tuxtuac$^5$                & 1975 October 16  & 30             & 24.5           & 2              &                            \\
          Parambu$^6$                & 1967 July 24     & 28             & 22.5           &                &                            \\
          Guidder                    & 1949 January 7   &                &                &                &                            \\
          Khanpur                    & 1932 July 8      &                &                &                &                            \\
          Konovo                     & 1931 May 26      &                &                &                &                            \\
          Paragould                  & 1930 February 17 &                &                &                &                            \\
          Olivenza$^{7, 8}$          & 1924 June 19     & 29.1           & 25             &                & 0.23$\pm$0.06              \\
          Ny\'{\i}r\'abr\'any$^{9}$  & 1914 July 17     & 26.7           & 20.5           &                &                            \\
       \hline
      \end{tabular}
      \label{meteorites}
     \end{table*}
%
%

%
%
     \begin{figure*}
        \centering
        \includegraphics[width=\linewidth]{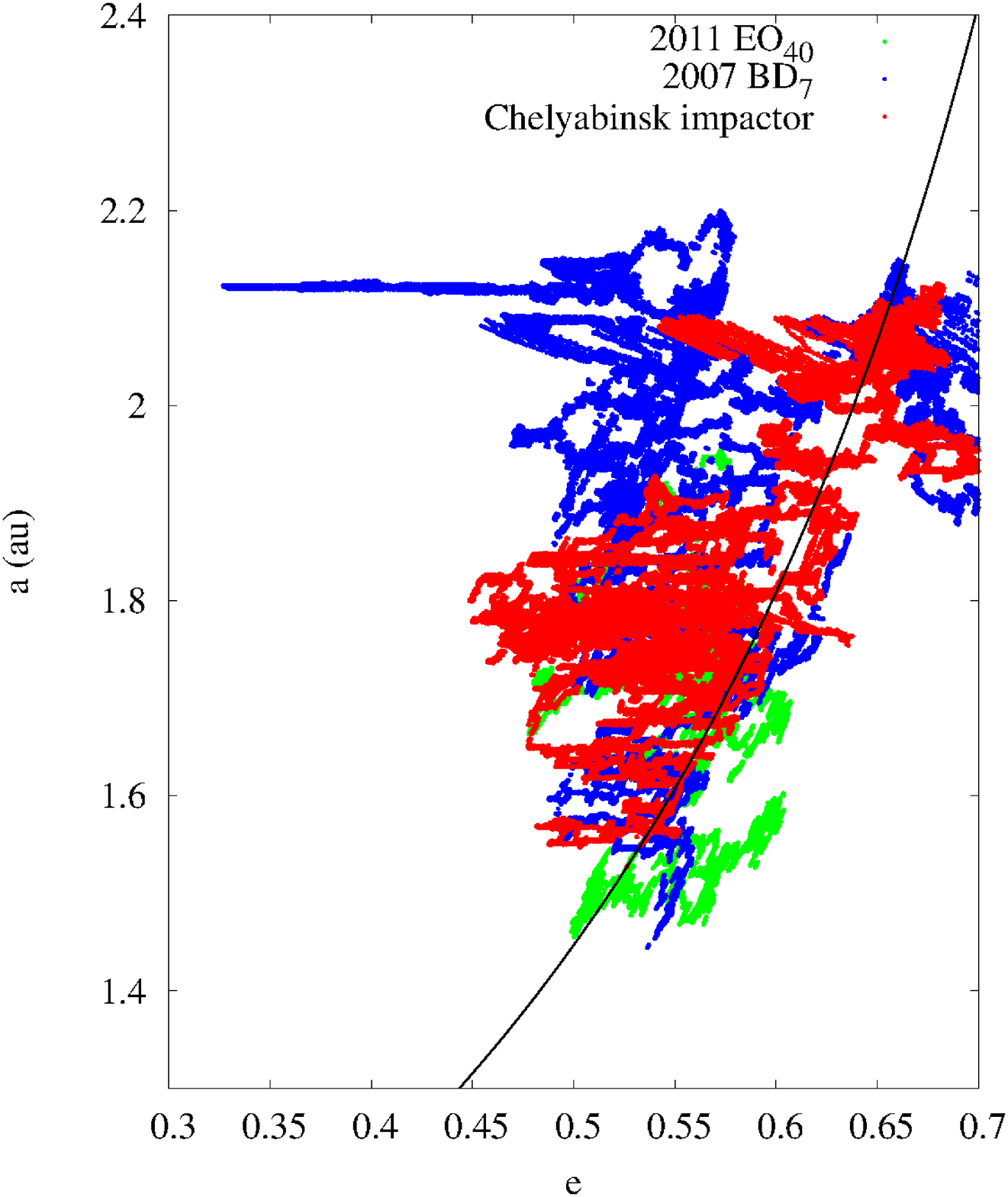}
        \caption{Backwards orbital evolution of the Chelyabinsk impactor, and asteroids 2007 BD$_{7}$ and 2011~EO$_{40}$. The continuous 
                 line represents the ($e$, $a$) combination with perihelion at the semimajor axis of Venus (0.7233 au). The objects 
                 experience multiple episodes of horizontal (resonant) oscillations as in Fig. 3, panel C of Gladman et al. (1996). Our 
                 calculations did not include the Yarkovsky effect which may have a non-negligible role on the medium, long-term evolution 
                 of objects as small as the ones studied here. Proper modeling of the Yarkovsky force requires knowledge on the physical 
                 properties of the objects involved (for example, rotation rate, albedo, bulk density, surface conductivity, emissivity) 
                 which is not the case for the objects discussed here. Detailed observations during future encounters with the Earth should 
                 be able to provide that information. On the short term, the Yarkovsky force mainly affects $a$ and $e$. Its effects are 
                 negligible if the objects are tumbling or in chaotic rotation. The non-inclusion of this effect has no major impact on the 
                 assessment completed. The time is referred to JDCT 2456200.5 = A.D. 2012-Sep-30 00:00:00 and we integrated the orbits for
                 0.7 Myr.
                 }
        \label{horizontal}
     \end{figure*}
%
%

\end{document}